\documentclass[
 reprint,
 amsmath,amssymb,
 aps,nofootinbib
]{revtex4-2}
\usepackage{graphicx,amsfonts,color,comment,amsmath,hyperref,float}
\usepackage[dvipsnames]{xcolor}
\usepackage{bigints}
\usepackage[normalem]{ulem}

\newcommand{\be}{\begin{equation}}
\newcommand{\ee}{\end{equation}}
\newcommand{\bs}{\begin{split}} 
\newcommand{\bea}{\begin{eqnarray}}
\newcommand{\eea}{\end{eqnarray}}

\begin{document}

\title{Quasithermal GeV neutrinos from neutron-loaded magnetized outflows in core-collapse supernovae: spectra and light curves}

\author{Jose A. Carpio$^{1,2}$, Nick Ekanger$^{3}$, Mukul Bhattacharya$^{1,4}$, Kohta Murase$^{1,5,6}$, Shunsaku Horiuchi$^{3,7}$} 
\affiliation{
${}^1$Department of Physics; Department of Astronomy \& Astrophysics; Center for Multimessenger Astrophysics,
Institute for Gravitation and the Cosmos, The Pennsylvania State University, University Park, PA 16802, USA\\
${}^2$Department of Physics \& Astronomy; Nevada Center for Astrophysics, University of Nevada, Las Vegas, NV 89154, USA\\
${}^3$Center for Neutrino Physics, Department of Physics, Virginia Tech, Blacksburg, VA 24061, USA\\
${}^4$Department of Physics, Wisconsin IceCube Particle Astrophysics Center, University of Wisconsin, Madison, WI 53703, USA\\
${}^5$Center for Gravitational Physics and Quantum Information, Yukawa Institute for Theoretical Physics,\\ Kyoto University, Kyoto, Kyoto 606-8502, Japan\\
${}^6$School of Natural Sciences, Institute for Advanced Study, Princeton, NJ 08540, USA\\
${}^7$Kavli IPMU (WPI), UTIAS, The University of Tokyo, Kashiwa, Chiba 277-8583, Japan 
}




\begin{abstract}
Rapidly rotating and strongly magnetized protoneutron stars (PNSs) created in core-collapse supernovae can drive relativistic magnetized winds. Ions and neutrons can be co-accelerated while they remain coupled through elastic collisions. We investigate the nucleosynthesis and subsequent nuclear disintegration, and find that relativistic neutrons can be generated in such magnetized winds. Upon eventual decoupling, resulting inelastic collisions with ejecta lead to pion production, resulting in $0.1-10\,{\rm GeV}$ neutrinos. Following this scenario presented in Murase, Dasgupta \& Thompson, Phys. Rev. D, 89, 043012 (2014), we numerically calculate the spectra and light curves of quasithermal neutrino emission. In the event of a Galactic supernova, $\sim 10-1000$ neutrino events could be detected with Hyper-Kamiokande, KM3Net-ORCA and IceCube-Upgrade for PNSs with surface magnetic field $B_{\rm dip}\sim 10^{13-15}\,{\rm G}$ and initial spin period $P_i \sim 1-30\,{\rm ms}$.
Successful detection will enable us to study supernovae as multienergy neutrino sources and may provide clues to the roles of PNSs in diverse classes of transients.
\end{abstract}


\maketitle

\section{Introduction}
Dying stars with masses $\gtrsim 8M_\odot$ generally result in core collapse, whereby the star releases almost all its gravitational binding energy in the form of $\sim 10\;{\rm MeV}$ neutrinos over a few tens of seconds. After the core collapse, if a supernova (SN) explosion occurs and a protoneutron star (PNS) is left as a remnant, neutrinos heat the surrounding material and drive a wind. 
As the PNS spins down, its rotation and magnetic energy would be transported outwards as Poynting flux which is further converted to kinetic energy of the outflow, such that the pulsar wind can be accelerated to relativistic speeds \cite{Michel1969,Lyubarsky:2000yn,Drenkhahn2002}. 
The PNS wind would be embedded deep inside ejecta, but if the spin-down power is large enough it may affect SN dynamics.
The PNS may be strongly magnetized and/or rapidly rotating, which may account for a variety of transients related to core-collapse SNe (CCSNe), including super-luminous SNe, hypernovae, and even gamma ray bursts (GRBs) \cite{Thompson2004,Woosley:2006fn,Bromberg:2012gp,Metzger:2015tra,Kashi2016,Margalit:2017oxz,Yu:2017fsg,Bhattacharya_2023}. 

The nuclear composition of the SN outflow largely consists of free protons and neutrons in the neutrino-heating phase following the core-collapse. 
In the early stages of magnetized outflows, the majority of these free nucleons may readily form seeds for nucleosynthesis or capture onto nuclei \cite{Vlasov_2017,Bhattacharya:2021cjc,Ekanger2022,reichert2022,Ekanger:2023mde}.
However, any nuclei that interact with ambient photons or matter can undergo photodisintegration or spallation, and release free neutrons (see e.g., Ref.~\cite{Horiuchi2012}). Free neutrons remain coupled to protons via elastic collisions during the early evolution phase when the density is high enough. However, as the density becomes sufficiently low due to the expansion, neutrons will decouple and be left behind by accelerating ion outflows. 

\begin{figure*}
\includegraphics[width=0.95\textwidth]{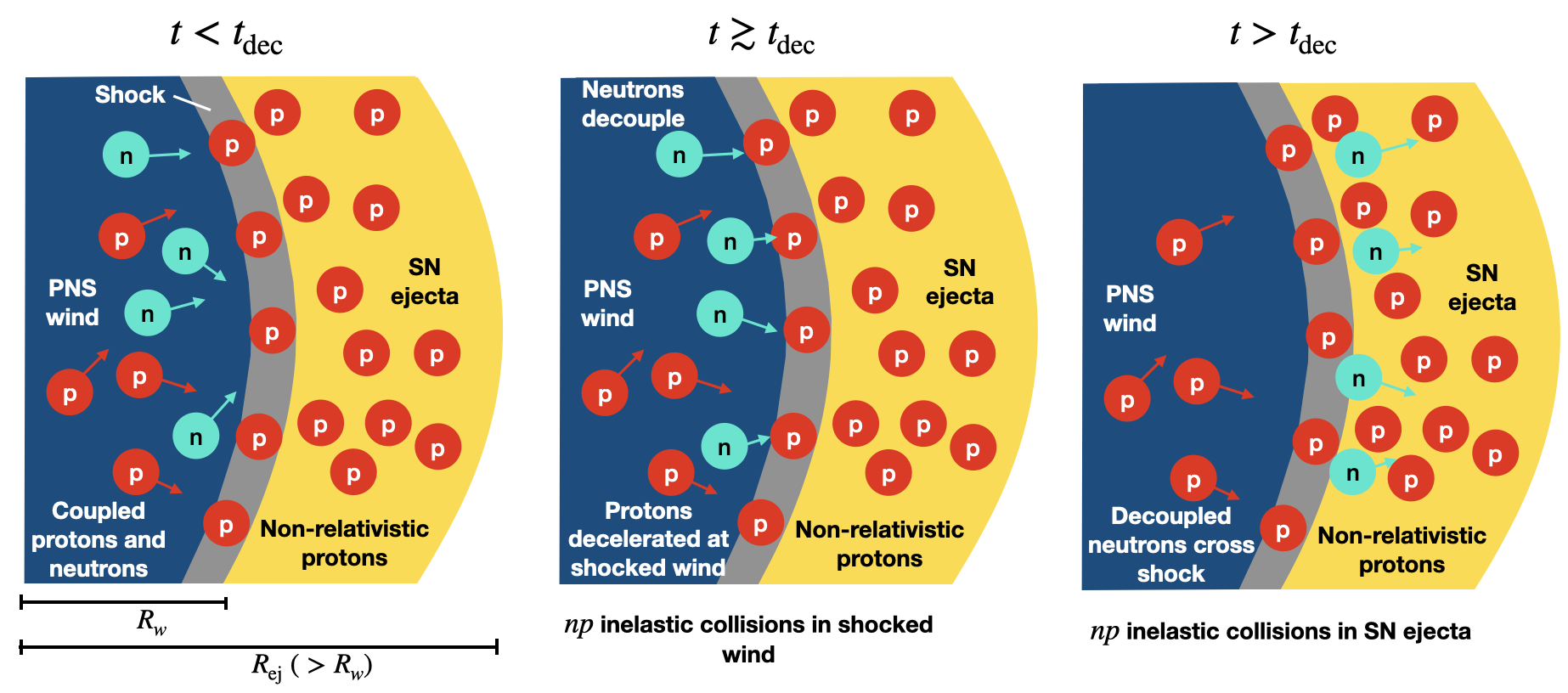}
\caption{Schematic diagram of the neutron-loaded PNS wind, with radius $R_w$. The SN ejecta is ahead of the wind ($R_{\rm ej} > R_w$) and contains nonrelativistic protons. Initially, before the neutron decoupling time, $t_{\rm dec}$, the neutrons accelerate with the protons via elastic $np$ coupling (left panel). After the decoupling, neutrons will move with a constant bulk velocity. The protons are promptly decelerated at the shock, while neutrons can be decelerated only via $np$ inelastic collisions in the shocked wind (middle panel).
At sufficiently late times, the neutrons can further enter the nonrelativistic SN ejecta, and $np$ interactions with protons in the ejecta occur (right panel).
}
\label{Figure1}
\end{figure*}

Murase, Dasgupta and Thompson (hereafter MDT14 \cite{MBT2014}) suggested that these neutrons eventually hit shocked nebulae and/or dense ejecta, and the relativistic neutrons exceeding the pion production threshold should initiate $np$ inelastic collisions and subsequently lead to quasithermal neutrinos with energies of $\sim 0.1 - 10\,{\rm GeV}$. The schematic picture of this scenario is explained in Fig.~\ref{Figure1}.
This mechanism does not require particle acceleration via shocks or magnetic turbulence, which is unlikely to be efficient in radiation-dominated environments \cite{MuraseIoka2013}. 

Detection of these neutrinos with upcoming detectors such as KM3NeT, Hyper-Kamiokande (HK) and IceCube-Upgrade would better inform the physics of magnetic outflows, nuclei synthesis and disintegration, in addition to the GRB-SN connection. 
The next Galactic CCSN will provide high-statistics observations of not only the thermal $\sim 10$~MeV neutrino burst \cite{OConnor:2018sti,Hyper-Kamiokande:2021frf}, but also the non-thermal TeV-PeV neutrino flare \cite{Murase:2017pfe}. 
Detecting neutrinos at GeV-TeV energies will fill the bridge and provide clues to physics of the outflows.
 
In this work, we consider the $\sim 0.1 - 1\,{\rm GeV}$ neutrinos produced from neutron-proton interactions in magnetized outflows originating from rapidly rotating PNSs. In Section~\ref{Sec2}, we discuss SESNe and the parameters that characterize the associated PNSs. 
In Section~\ref{Sec3}, we discuss the properties of neutrino-driven winds from PNSs. In Section~\ref{Sec4}, we evaluate the criteria for synthesized nuclei to disintegrate within the magnetized outflow and estimate the fraction of free neutrons. In Section~\ref{Sec5}, we look at neutrino emission from choked outflows and their detection prospects for KM3Net-ORCA, IceCube-Upgrade and HK. Finally, we discuss our results in Section~\ref{Sec6} and conclude in Section~\ref{Sec7}.

\section{Pulsar-aided supernovae} 
\label{Sec2}
We explore a wide parameter range: initial spin period $P_i \sim 1-30\, {\rm ms}$, surface dipolar magnetic field $B_{\rm dip} \sim 10^{13}- 10^{15}\, {\rm G}$; corresponding to SN ejecta mass $M_{\rm ej} \sim 1-10\, M_{\odot}$ and explosion energy ${\mathcal E} \sim 10^{51}-10^{52}\, {\rm erg}$. 
Pulsars can be harbored in Type II SNe, as seen in Galactic pulsars such as the Crab pulsar \cite{Smith:2013gya}. The Crab pulsar has a present spin period of $33\,{\rm ms}$ and an estimated initial spin period of $\sim 16-19\,{\rm ms}$ \cite{Kou:2015oma}. 

Strongly magnetized and rapidly spinning PNSs are contenders for the central engines of both long-duration GRBs and SLSNe.
They can be left as remnants of SESNe, including SNe Ibc and SNe Ibc-BL. A long-lived central engine explains a hydrogen-poor class of SLSNe, which are stellar explosions with peak luminosities $\sim 10-100$ times higher than normal CCSNe \citep{Quimby2011,GalYam2012}. In engine-powered SLSN-I, the central compact object is believed to be either a millisecond PNS \citep{Maeda:2007ck,Woosley2010} or a BH with an accretion disk \citep{DK2013}. 
 
The formation rate of these PNSs would be lower than the CCSN rate $\mathcal{R}_{\rm CCSN}$, which we assume is equal to the Galactic CCSN rate of $3.2^{+7.6}_{-2.6}$ per century per galaxy (see, e.g., Ref.~\cite{Adams_2013}). 
The observed event rate of high-energy transients associated with SESNe is relatively small (see, e.g., Ref.~\cite{Kashi2016}): the low-luminosity GRB rate is $\sim 0.01\times\mathcal{R}_{\rm CCSN}$ and for long GRBs, assuming a beaming factor of $\sim 100$, the rate is $\sim  10^{-4}\times\mathcal{R}_{\rm CCSN}$\cite{Kashi2016}.

\section{Neutrino-driven winds and neutron decoupling}
\label{Sec3}
Neutrinos heat the PNS surface to generate a baryonic wind with a mass-loss rate~\cite{QW1996, Metzger2011a}
\begin{equation}\label{Mdot}
\Dot{M}_b\approx 5\times10^{-5}~\textrm{M}_\odot~\textrm{s}^{-1}~\mathcal{F}_{\rm mag}C_{\rm inel}^{5/3}L_{\nu,52}^{5/3}\varepsilon_{\nu,10}^{10/3}R_{10}^{5/3}M_{\rm NS, 1.4}^{-2},
\end{equation}
where $\mathcal{F}_{\rm mag}=f_{\rm op}f_{\rm cen}$.
Here $f_{\rm op}$ denotes the fraction of the PNS surface threaded by open magnetic field lines, $f_{\rm cen}$ is the magnetocentrifugal enhancement factor, $C_{\rm inel}$ is a correction factor for inelastic neutrino-electron scatterings, $L_{\nu}=L_{\nu,52}\times10^{52}\, {\rm erg\, s^{-1}}$ is the $\nu_e+\bar{\nu}_e$ neutrino luminosity, 
$\varepsilon_{\nu}=\varepsilon_{\nu,10}\times10\, {\rm MeV}$ is mean neutrino energy, $R_{\rm NS}=R_{\rm NS, 10}\times10\, {\rm km}$ is PNS radius and $M_{\rm NS}=M_{\rm NS, 1.4}\times1.4\, M_{\odot}$ is PNS mass. 
For our analysis, we adopt the neutrino quantities for a NS with mass $M_{\rm NS}=1.4\, M_{\odot}$ and radius $R_{\rm NS}=10\,{\rm km}$ from Ref.~\cite{Pons1999}. We also incorporate a stretch factor $\eta_s=3$ to account for a longer PNS cooling timescale due to its rapid rotation~\cite{Metzger2011a}. 

The neutrino-thin time, $t_{\rm thin}$, is the time when the PNS becomes transparent to SN neutrinos in the MeV range. Consequently, once $t \gtrsim t_{\rm thin}$, $L_\nu$, $\varepsilon_{\nu}$ and $\dot{M}_b$ all decline more rapidly than the initial power-law decrease. The precise value of $t_{\rm thin} \approx 30\,{\rm s} - 100\,{\rm s}$ is sensitive to the properties of neutrino opacities, which depends on the unknown NS equation of state as well as the NS rotation rate~\citep{Metzger2011a}. 
This time scale gives a reasonable proxy for the time window $\Delta T$ in the  signal search with neutrino detectors. In this work, we assume $\Delta T=t_{\rm thin}=70$~s. Note that we calculate neutrinos after $t_{\rm thin}$ even if we do not expect significant contributions to the neutrino signal after $t_{\rm thin}$. The impacts of different choices of $\Delta T$ are discussed in Section VI.

\begin{figure}
    \centering
\includegraphics[width=1.02\columnwidth]{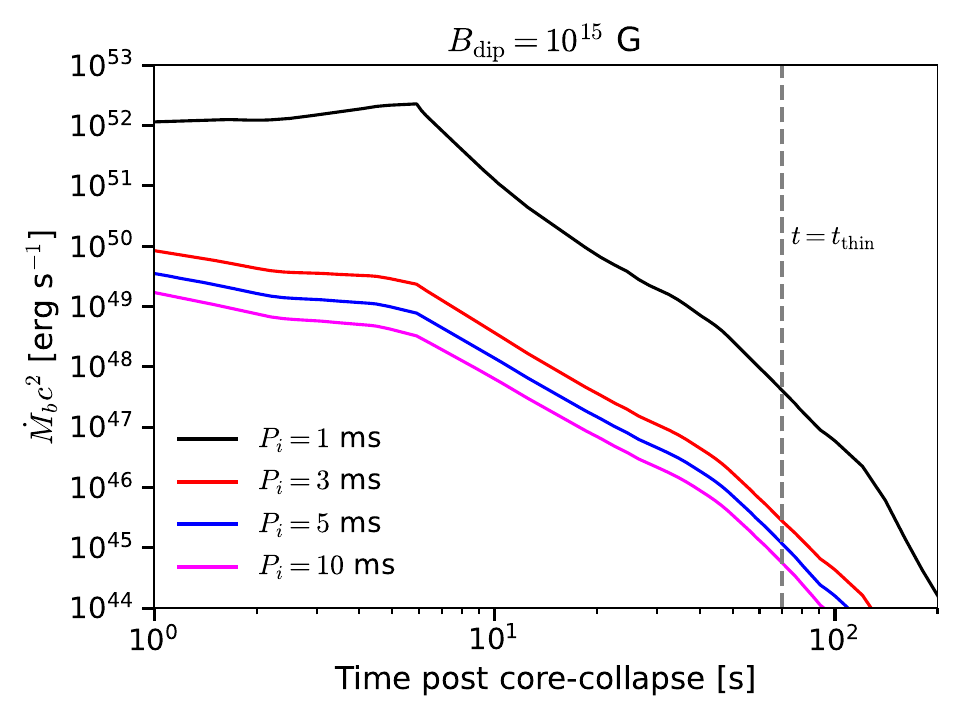}
    \vspace{-0.6cm}
    \caption{Mass-loss rate scaled by $c^2$, $\dot{M}_bc^2$, as a function of time, at $B_{\rm dip}=10^{15}$ G and different spin periods $P_i$. The gray line corresponds to  $t=t_{\rm thin} = 70\;{\rm s}$.}
    \label{MdotPlot}
\end{figure}

We show in Fig.~\ref{MdotPlot} the time dependence of $\Dot{M}_b$ for different configurations. The shaded region corresponds to times $t>t_{\rm thin}$, where $\dot{M}_b$ drops by several orders of magnitude. In our calculations, contributions to the neutrino flux are essentially negligible beyond $t_{\rm thin}$ (see Section V). Throughout the first $\sim 6$~s, the PNS is still contracting, which in turn increases the PNS angular velocity and $\dot{M}_b$. The differences in $\dot{M}_b$ for different PNS parameter sets are set by the $B_{\rm dip}$ and $P_i$ dependence in $\mathcal{F}_{\rm mag}$.

As the wind expands radially outwards, the Lorentz factor $\Gamma$ evolves as \cite{Drenkhahn2002}
\begin{equation}
\label{LorentzFactor}
\beta\Gamma=
\begin{cases} 
\sigma_0(r/R_{\textrm{mag}})^{1/3}, & r\leq R_{\textrm{mag}} \\
      \sigma_0, & r>R_{\textrm{mag}} 
\end{cases},
\end{equation}
where $r$ is the distance of the outflow from the PNS and $R_{\rm mag}$ is the magnetic dissipation radius, given by
\begin{equation}
\label{R_mag}
    R_{\rm mag}\approx(5.0\times10^{12}\, {\rm cm})\left(\frac{\sigma_0}{10^2}\right)^2\left(\frac{P}{\rm ms}\right)\left(\frac{\epsilon}{0.01}\right)^{-1}.
\end{equation}
Here $\epsilon \sim 0.01$ is a parameter that is used to describe the reconnection velocity.

The magnetization is $\sigma_0=\phi_B^2\Omega^2/\Dot{M}c^3$, where $\phi_B=(f_{\rm op}/4\pi)B_{\rm dip}R_{\rm NS}^2$ is the magnetic flux due to a rotating dipole field with magnitude $B_{\rm dip}$ and $\Omega$ is the PNS angular velocity. 

Above $R_{\rm mag}$, the Poynting flux has dissipated and been converted to kinetic energy, saturating the outflow Lorentz factor to $\sigma_0$. The outflow velocity quickly becomes relativistic, with a weak dependence on the initial outflow conditions $\sigma_0$ and $R_{\rm mag}$. We will consider $10^{13}\; {\rm G} \leq B_{\rm dip}\leq 10^{16}\;{\rm G}$ and $1\; {\rm ms}\leq P_i \leq 30\;  {\rm ms}$. When $\sigma_0 \gtrsim 1$, the outflow becomes relativistic.

The wind, which pushes the ejecta including the cocoon material, is significantly decelerated after the wind termination shock, forming a hot magnetized bubble. The evolution of the cavity and nebula depends on the PNS spin-down power and the ejecta \citep[e.g.,][]{CF1992,Bucci2009,Kotera2013}. Magnetic dissipation inside the tenuous wind bubble with radius $R_w$ has been considered in the context of magnetar models for SNe Ibc-BL, SLSNe and rapidly rising optical transients \cite{Kashi2016,Hotokez2017,Margalit:2017oxz,Margalit:2018bje}. 

To compute the wind radius $R_w$ and ejecta radius $R_{\rm ej}$, we numerically solve the differential equations \cite{Murase:2014bfa,Kashi2016}
\begin{eqnarray}
\frac{dR_w}{dt}&=&\sqrt{\frac{7}{6(3-\delta)}\frac{\mathcal{E}_{\rm tot}}{M_{\rm ej}}\left(\frac{R_w}{R_{\rm ej}}\right)^{3-\delta}}+\frac{R_w}{t}\\
\frac{dR_{\rm ej}}{dt}& \equiv &V_{\rm ej} = \sqrt{\frac{2\mathcal{E}_{\rm tot}}{M_{\rm ej}}}.
\label{DynamicsEq}
\end{eqnarray}
where $\delta=1$ is used for the ejecta density profile. We assume the ejecta mass $M_{\rm ej}=3M_\odot$ and ejecta energy $\mathcal{E}_{\rm tot}=10^{51}\,{\rm erg}$.
The first term in the right-hand side of Eq.~\eqref{DynamicsEq} is the expansion velocity. The initial condition for the ejecta and wind radii is $R_w(t=0) = R_{\rm ej}(t=0) = R_{\rm LC} = cP_i/2\pi$, where $R_{\rm LC}$ is the light cylinder radius. Note that the evolution close to the light cylinder radius is different (see, e.g., Ref.~\cite{Michel1969}) but our results are insensitive to the choice of the initial radii.

Neutrons and ions have the same outflow velocity as
long as they are coupled together with $\langle \sigma_{\rm el}v\rangle \approx \sigma_{\rm np} c$. 
Here $\sigma_{\rm el}$ is the elastic cross section and $\sigma_{np}\approx 3\times10^{-26}\, {\rm cm}^2$ is
the inelastic cross section. The nucleon number density in the wind is $n_w = \dot{M}_b/4\pi R_w^2 cm_p \Gamma_w$ and the optical depth for $np$ collisions is $\tau_{np} = n_w \sigma_{\rm np}(R_w/\Gamma_w)$. The decoupling radius $R_{\rm dec}$ where $\tau_{np}=1$ is given by \cite{MBT2014}
\begin{equation}
R_{\rm dec}\approx (2.5\times10^9\, {\rm cm})\sigma_{0,3}^{-1} B_{\rm dip,15}^{6/5} P_{i,-3}^{-4/5}f_{\rm op,-1}^{6/5}\epsilon_{-2}^{-2/5}
\end{equation}
and beyond this radius the neutrons are decoupled and no longer experience bulk acceleration.
Typically, $R_{\rm dec}$ exceeds both $R_{\rm mag}$ and $R_{\rm w}$ at early times. It gradually decreases over time while $R_{\rm mag}$ and $R_{\rm w}$ both increase.

\section{Nuclear disintegration}\label{disintegration}
\label{Sec4}
The fraction of free neutrons in the outflow, $Y_n$, depends on both the ability to synthesize nuclei and the photodisintegration efficiency. At early times post-bounce, the density and temperature in the outflow are favorable for nuclei to be synthesized, but the outflow will not synthesize nuclei all the way until $t_{\rm thin}$. Even at early times when most neutrons are used for nucleosynthesis, if the nuclei are disintegrated by, e.g., ambient photons, the amount of free neutrons can be restored.

The photons can arise from both thermal and nonthermal processes, the relative importance of which are determined by the interaction optical depth at the radius of the termination shock. While thermal photons have typical energies of $\sim 10\,{\rm keV}$ and may not be energetic enough to efficiently initiate photodisintegration, nonthermal photons can reach energies of $\sim 1\,{\rm MeV}$ through synchrotron cascades which allows for photodisintegration \cite{MBT2014}. In this section, we discuss the ways in which the outflow may achieve high free neutron fractions, covering the survivability of nuclei against the photon fields at the termination shock that leak into the unshocked wind. In later sections, we assume that any nuclei synthesized are disintegrated into free neutrons and protons.

\subsection{Photodisintegration}

Nuclei in the unshocked wind interact with photons that are produced in the wind via dissipation and photons that have leaked in from the nebula or shocked wind. In this work, we consider the former, where the dissipation can occur in the current sheet \cite{2014ApJ...780....3U,Cerutti:2014ysa,Philippov:2017ikm}. 
The Thomson optical depth in the wind, $\tau_T$, is written as $\tau_T \approx y_\pm n_w\sigma_TR_w/\Gamma_w$, where $\sigma_T$ is the Thomson cross section and $y_\pm$ is an enhancement factor that accounts for pair production in the wind \cite{MBT2014}. The enhancement factor is defined as $y_\pm = (n_w+n_{\pm})/n_w$, where $n_{\pm}=\dot{M}_{\pm}/(4\pi R_w^2 c m_e\Gamma_w)$ is the pair number density, $\dot{M}_{\pm}\approx (2.5\times 10^{-17}\,M_\odot/{\rm s})\, \mu_{\pm,6}B_{\rm dip,15}P_{i,-2}^{-2}R_{\rm ns,6}^3$ is the Goldreich-Julian (GJ) density \cite{1969ApJ...157..869G} and $\mu_\pm\sim 10^5-10^7$ is the pair multiplicity, the number of pairs produced by each primary accelerated particle. The pair multiplicity value can vary within $10^5-10^7$, and we take $\mu_\pm=10^6$ as the fiducial value \cite{2011MNRAS.410..381B,Timokhin:2015dua}. At early times $y_\pm \sim 1$, but it grows when $\dot{M}_b$ becomes comparable to $\dot{M}_{
\pm}$. 

If $\tau_T \gg 1$, the photons get largely thermalized to a temperature $T_\gamma$ in the comoving frame of the nebula, which satisfies $aT_\gamma^4 = \epsilon_{\gamma}\Gamma_w(\dot{M}_b+\dot{M}_{\pm})c^2/(4\pi \Gamma_w^2 R_w^2 c)$, where $\epsilon_\gamma=0.3$ is the fraction of the total energy as radiation energy. At early times, we find that $\tau_T\gtrsim1$ for all $B_{\rm dip}$ and $P_i$ configurations considered here, but $\tau_T$ reduces to close to unity at $t_{\rm thin}$ for longer spin periods, e.g., $P_i=30\,{\rm ms}$.

To estimate the effect of photodisintegration on nuclei, we calculate the efficiency as (see also Ref.~\cite{MBT2014}):
\begin{equation}\label{efficiency}
f_{A\gamma}\approx \kappa_A \sigma_{A\gamma} n_{\gamma}(R_w/\Gamma_w),
\end{equation}
where $\kappa_A$ is the nuclei inelasticity, $\sigma_{A\gamma}$ is the photodisintegration cross section and $n_{\gamma}$ is the photon number density. 
For nuclei in our magnetized winds, the dominant photodisintegration contribution comes from the giant dipole resonance (GDR), which occurs predominantly at energies comparable to the nuclear binding energy $\sim 10\,{\rm MeV}$ in the nuclei rest frame. We estimate $\kappa_A\sigma_{A\gamma}\sim$ ${\rm 1.4\times10^{-27}\,cm^2}$ $(A/56)^{1/6}$ \cite{1996PhDT........59R}, where we take $A=56$ as a representative value for nuclei that could be synthesized in these winds.

Since we find $\tau_T\gtrsim 1$ for all configurations at early times ($t \lesssim t_{\rm thin}$), we expect that the ambient photons follow a thermal distribution with 
$n_{\gamma}\approx19.23(k_B T_{\gamma}/hc)^3$ \cite{Bhattacharya:2021cjc}. The thermal photon temperature is greatest at early times post-bounce and is around $2\times10^9\,{\rm K}\sim200\,{\rm keV}$ for $P_i=1\,{\rm ms}$ and $4\times10^8\,{\rm K}\sim40\,{\rm keV}$ for $P_i=30\,{\rm ms}$ (the temperature depends mostly on the spin period and is roughly independent of $B_{\rm dip}$). The temperature then decreases monotonically with time and distance from the light cylinder. Since the photon temperatures are $\sim 10-100\,{\rm keV}$, the threshold GDR temperature ($k_B T_{\gamma}<\varepsilon_{\rm th}\sim10\,{\rm MeV}$) is not reached, and we find that the nuclei can survive. However, when these temperatures are close to the threshold, a more careful numerical integration of the efficiency is warranted which we will consider in a future work (Ekanger et al. 2024, in prep).

Because $\tau_T$ is of order unity for some configurations (especially those with higher $P_i$ values and at later times near $t_{\rm thin}$), and our calculation of $\tau_T$ is model dependent, there could still be leakage of nonthermal photons in the outflow. For nonthermal photons the number density of photons above energy $\varepsilon_{\rm th}$ is  $n_{\gamma}\sim (\epsilon_e\Gamma_w(\dot{M}_b+\dot{M}_{\pm})c^2/[4\pi \Gamma_w^2 R_w^2 c\ \varepsilon_{\rm syn}])(\varepsilon_{\rm th}/\varepsilon_{\rm syn})^{-0.5}$, where $\epsilon_e \sim 0.1$ is the fraction of energy in electrons, $\varepsilon_{\rm syn}$ is the characteristic synchrotron energy. This approximation is valid for photons in the synchrotron fast cooling regime. Nonthermal photons may lead to $f_{A\gamma}\gg1$ at early times for $R_w \sim R_{\rm LC}$ \cite{MBT2014}, suggesting that, if there is a significant component of nonthermal photons, the nuclei could be easily disintegrated. 

\subsection{Other disintegration processes}

\begin{figure*}
\includegraphics[width=0.49\textwidth]{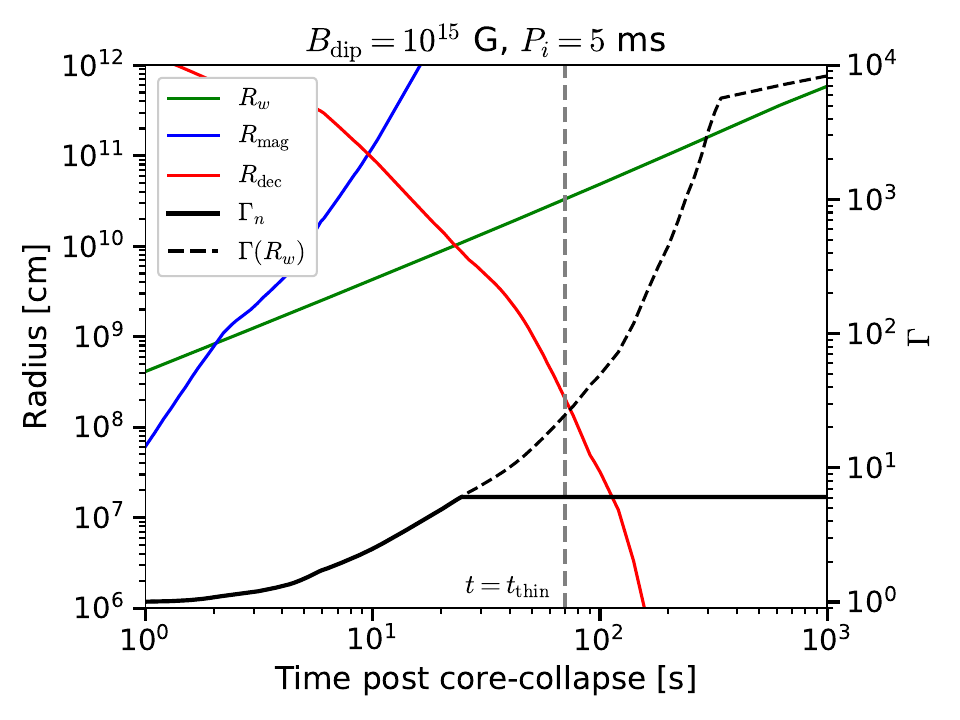}
\includegraphics[width=0.49\textwidth]{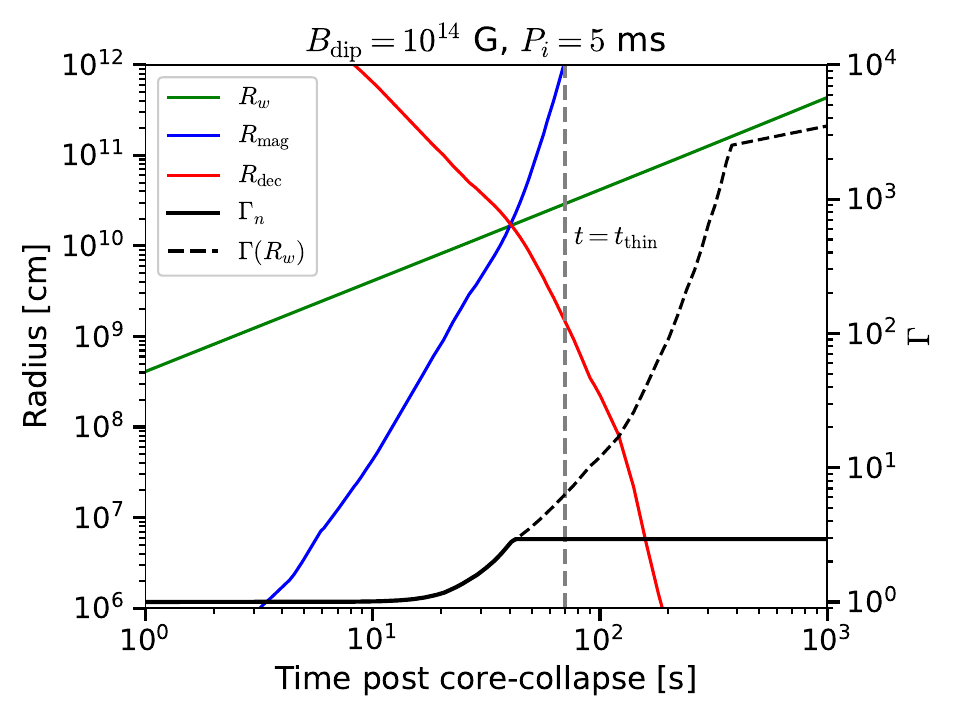}
\caption{Evolution of $R_w,R_{\rm mag}$ and $R_{\rm dec}$, shown as blue, green and red curves, respectively. We also show $\Gamma_w$ and $\Gamma_n$ as solid and dashed black lines, respectively. The neutron Lorentz factor saturates to its terminal value after $R_w = R_{\rm dec}$ and neutrons subsequently decouple from the outflow. The gray vertical line marks $t=t_{\rm thin} = 70\;{\rm s}$. Left (right) panel corresponds to the PNS parameters $B_{\rm dip}=10^{15}$~G ($10^{14}$~G) and $P_i=5$~ms.}
\label{REvolution}
\end{figure*}

As discussed in MDT14, neutrons can be produced through nuclear disintegration by photons generated in the nebula. Leaked photons from the shocked wind are boosted in the unshocked wind. When $\tau_T\gg1$, only a fraction $f_{\rm esc}\sim 1/\tau_T$ of thermal photons will ultimately leak into the unshocked wind. In the case where $\tau_T<1$, photons are not thermalized and due to low optical depth, the nonthermal photons leak into the outflow with a leakage fraction $f_{\rm esc}\sim 1$. Nonthermal photons from the magnetized wind nebula \cite{Murase:2014bfa,Murase:2016sqo} may have broadband spectra and they can be energetic enough to lead to photodisintegration of the heavy nuclei. Even if nuclei mostly survive, neutrons could still be produced through photodisintegration in this nebula. As heavier elements are charged, their Lorentz factors would be equal to $\Gamma_w$ as they accelerate together with protons. The decoupling radius is expected to increase, as $n_w$ would decrease by a factor of $A$ while $\sigma_{Ap}$ scales as $A^{2/3}$. In the shocked wind, the nuclei have an average Lorentz factor of $\sim (\Gamma_w-1)$. Depending on nebular photon spectra, $f_{A\gamma} \gtrsim 1$ is possible close to the termination shock region, leading to neutron production.

Further, in addition to GDR, other photonuclear reactions can also photodisintegrate heavy nuclei. For higher photon energies in the nuclei rest frame, quasideuteron, pion production, and fragmentation processes become relevant. 
However, the thermal photons in our system do not typically reach such high energies. Finally, GDR successfully photodisintegrates nuclei that are synthesized until the point in time when the particle density is too low such that additional nucleosynthesis cannot occur. 

Synthesized nuclei may also undergo spallation. To avoid spallation due to nucleon-nucleon collisions, the relative velocity between the outflow and the surrounding material should not exceed a critical value $\beta_{\rm sp} \approx 0.14$, at which the relative kinetic energy equals the nuclear binding energy $\sim10\,{\rm MeV}$. The spallation rate due to collision with other nucleons in the wind can be estimated as, $t_{\rm sp}^{-1} = \sigma_{\rm sp}n_N \beta c$ (see, e.g, Ref.~\cite{Wang:2007xj}). Here, $\sigma_{\rm sp} = 5\times10^{-26}A^{2/3}\, {\rm cm^2}$ is the cross section for spallation of a nucleus with atomic number $A$, $n_N$ is the number density of nuclei in the sources, and $\beta$ is the velocity of the nucleons. However, spallation is not important for nuclei in our system as they are never energetic enough for their relative kinetic energy to exceed the nuclear binding energy. 

In this work, we calculate the photodisintegration efficiency with the analytical estimate given by Eq.~\ref{efficiency}. However, a more extensive analysis in the context of nuclei survival in different outflow environments and taking into account higher photon energy reactions, will be presented in a future work (Ekanger et al. 2024, in prep). This analysis will also be used to infer the parameter space for NS magnetic field and spin period where the nuclei can be synthesized and subsequently survive disintegration.

\section{Neutrino production and detection prospects}
\label{Sec5}

For the production of quasithermal neutrinos, we adopt the setup outlined in MDT14. Free neutrons in the outflow will couple to ions via elastic neutron-proton scatterings. Protons and neutrons will gain energy through bulk acceleration. Pion decay then leads to the production of $0.1 - 10\,{\rm GeV}$ neutrinos. In our study, the free neutrons are assumed to come from the photodisintegration of heavy nuclei synthesized. Throughout this section, we will adopt $Y_e = 0.5$ which yields a constant $Y_n = 0.5$, since we assume nuclei are fully photodisintegrated. 

Neutrons generated due to photodisintegration of heavy nuclei couple with ions via elastic neutron-proton scatterings. Through this coupling, the neutron Lorentz factor matches that of the outflow, $\Gamma_n = \Gamma_w$. Neutrons remain coupled to the outflow provided $\tau_{np} = n_w \sigma_{np}R/\Gamma_w>1$, with $n_w=\dot{M}_b/(4\pi R_w^2 m_p c \Gamma_w)$. The neutrons eventually decouple at the decoupling radius $R_{\rm dec}$, defined by $\tau_{np}(R_{\rm dec}) = 1$. We find that $R_{\rm dec}\propto \sigma_0^{-1}$, so it will decrease with time. On the other hand, $R_{\rm mag}\propto\sigma_0^2$ and $R_w\propto t$, where $t$ is the time post core-collapse. Hence, both $R_w$ and $R_{\rm mag}$ will eventually exceed $R_{\rm dec}$.

The time dependence of the relevant radii and Lorentz factors is shown in Fig.~\ref{REvolution} (cf. Fig.~1 of MDT14). The left panel depicts the case $B_{\rm dip}=10^{15}\,{\rm G}$ and $P_i=5$ ms which results in $R_{\rm mag}>R_{\rm dec}$ achieved before $R_w>R_{\rm dec}$. Neutrons have the Lorentz factor $\Gamma_w$ until the decoupling time when $R_w>R_{\rm dec}$ is reached. After this decoupling time, the neutrons achieve their maximum Lorentz factor $\Gamma(R_{\rm dec})$. The right panel shows the evolution of radii for $B_{\rm dip}=10^{14}\,{\rm G}$ and $P_i=5$ ms, where the radii intersect at roughly the same time. Note that both panels show cases where neutrons decouple before $t_{\rm thin}$ and the neutrons achieve $\Gamma_n>1.37$, allowing for pion production.

\subsection{Quasithermal neutrino emission}\label{neutrinoemission}

\begin{figure*}
\includegraphics[width=0.49\textwidth]{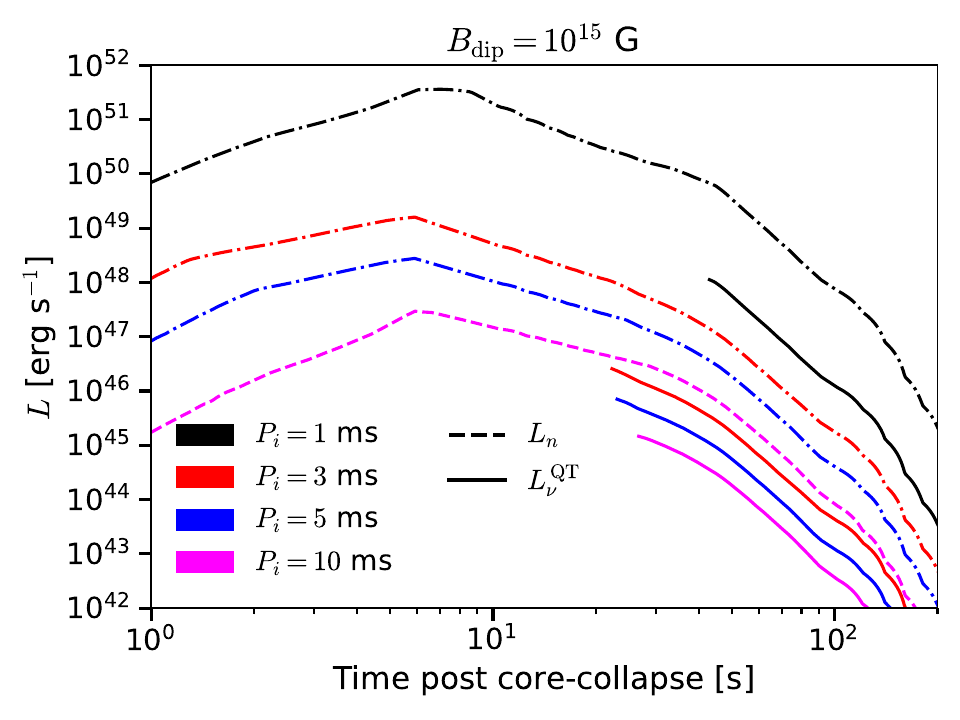}
\includegraphics[width=0.49\textwidth]{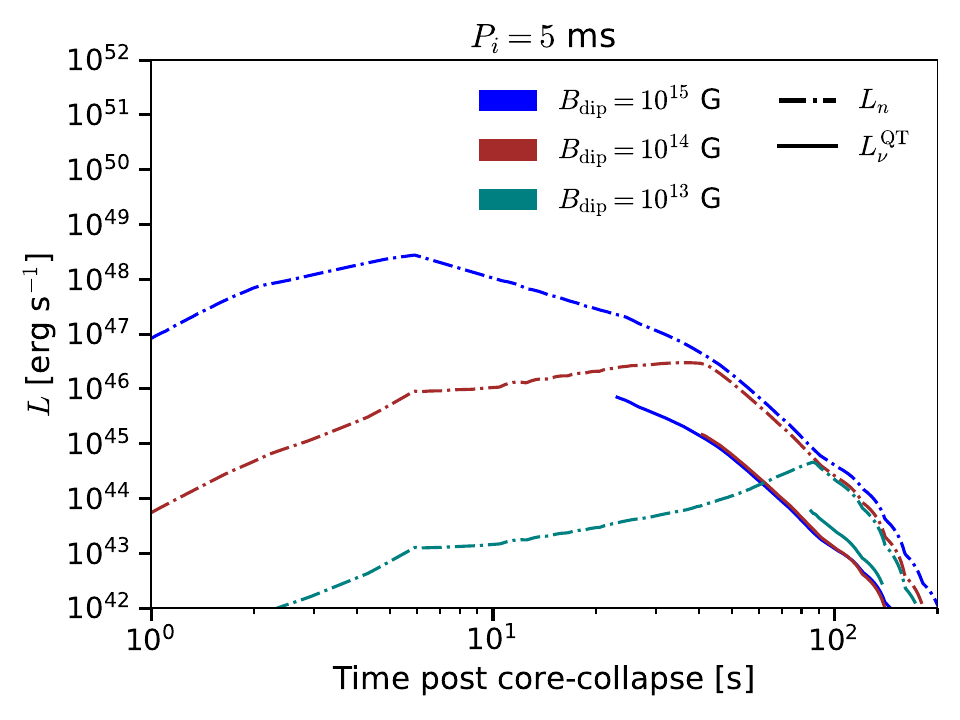}
\caption{Luminosity $L$ as a function of time post core-collapse, for $Y_n=0.5$. Left (right) panel assumes fixed $B_{\rm dip}=10^{15}\,{\rm G}$ ($P_i = 5\,{\rm ms}$). Solid (dot-dashed) lines correspond to quasithermal neutrino (neutron) luminosities. Neutrino emission is not expected until the neutrons are decoupled and have reached the pion production threshold. Note that in the right panel, the neutrino luminosity for $B_{\rm dip}=10^{15}\,{\rm G}$ and $B_{\rm dip}=10^{14}\,{\rm G}$ overlap with each other, the latter achieving the neutron decoupling at $t\gtrsim 40\;$s. 
}
\label{NeutronLuminosity}
\end{figure*}

If we consider the configuration with $B_{\rm dip}<10^{14}\,{\rm G}$ and $P_i>5\,{\rm ms}$, we may encounter cases where $R_w>R_{\rm dec}$ occurs first. In those situations, we define the time when $R_w=R_{\rm mag}$ as the neutron decoupling time. Before this decoupling time, we have $\Gamma_n=\Gamma(R_{\rm mag})$, instead of $\Gamma(R_w)$. The maximum Lorentz factor  remains as $\Gamma_n=\Gamma(R_{\rm dec})$ after decoupling time. One common feature of this scenario is that $\sigma_0$ takes time to build up, which is also why $R_{\rm mag}$ takes longer to exceed $R_{\rm dec}$. As a result, it takes $\gtrsim 10\,{\rm s}$ for the outflow to become relativistic and reach the pion production threshold.

At a given time $t$, the accelerated neutrons have energy $\varepsilon_n = \Gamma_n(t) m_nc^2$ and luminosity $L_n= Y_n\Gamma_n\dot{M}_b c^2$. 
The neutron luminosity $L_n$ (dot-dashed lines) and quasitermal neutrino luminosity $L_{\nu}^{\rm QT}$ (solid lines)is shown as a function of time in Fig. \ref{NeutronLuminosity}. We see that $L_n$ peaks at $\sim 6$ s for several $B_{\rm dip}$ and $P_i$ configurations. Note that $B_{\rm dip}=10^{15}\,{\rm G}$ and $P_i=1\,{\rm ms}$ has a neutron luminosity that peaks at $\sim 10^{52}\,{\rm erg~s}^{-1}$, which is extremely large. This is in part driven by $f_{\rm cen}\gtrsim 10^2$ for rapidly rotating PNSs with $P_i \lesssim 1.5\,{\rm ms}$ \cite{Metzger2011a}. As PNSs with such rapid spin periods are very rare \cite{ATNF}, we choose our representative case to be $P_i=5$~ms. For $P_i=5$~ms and $B_{\rm dip}=10^{13}$~G, $L_\nu^{\rm QT}$ essentially drops to zero at $t\sim 200$~s, because $\Gamma_n=\Gamma(R_{\rm dec})=1.45$ at the time of decoupling is just above pion production threshold. Over time, the neutron Lorentz factor decreases slightly and, in this case, drops below the threshold.

After $t\sim 6\,{\rm s}$, $\dot{M}_b$ drops as an approximate power law. Once the neutrons become relativistic, the Lorentz factor also exhibits an approximate power law scaling, until it achieves the maximum value at decoupling. These two scalings lead to the neutron luminosity dropping as a power law between 6 seconds until approximately 40 s, when $\dot{M}$ loses its power law behavior. For example $B_{\rm dip}=10^{15}\,{\rm G}$ and $P_i=1\,{\rm ms}$ has $L_n\propto t^{-2.3}$, but the index depends on the PNS parameters. As we decrease $B_{\rm dip}$ and the wind takes longer to become relativistic, the neutron's kinetic luminosity becomes suppressed by the factor $\sim(\Gamma_n-1)$. While the wind is non-relativistic, the factor $(\Gamma_n-1)$ and neutron luminosity $L_n$ do not behave as power laws, as seen in Fig. \ref{NeutronLuminosity} for $B_{\rm dip}=10^{13}\,{\rm G}$.

\begin{figure*}
\includegraphics[width=0.49\textwidth]{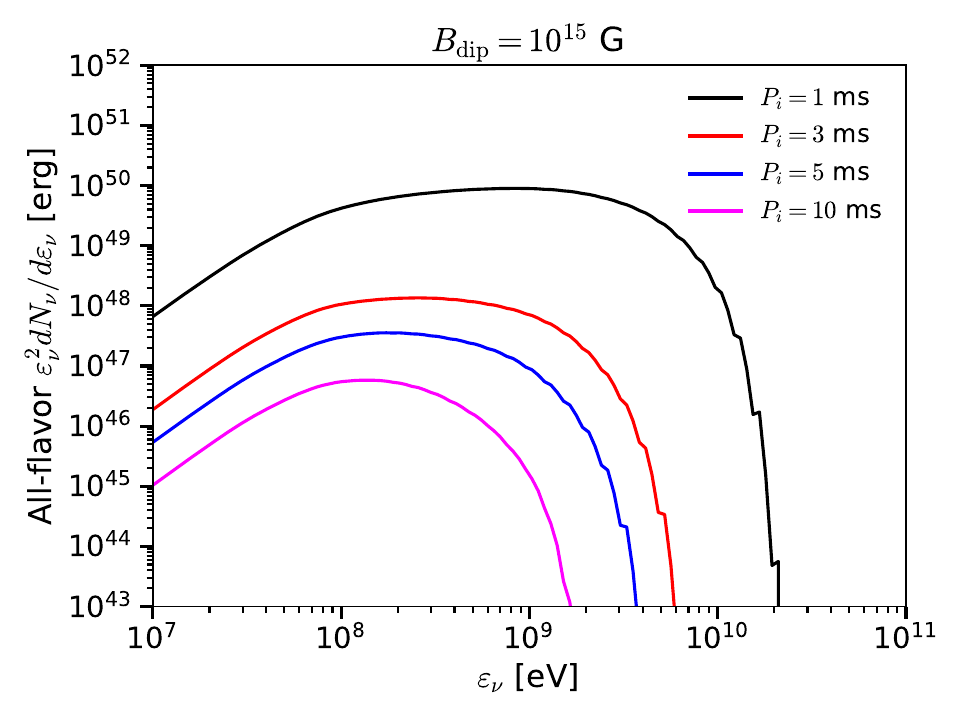}
\includegraphics[width=0.49\textwidth]{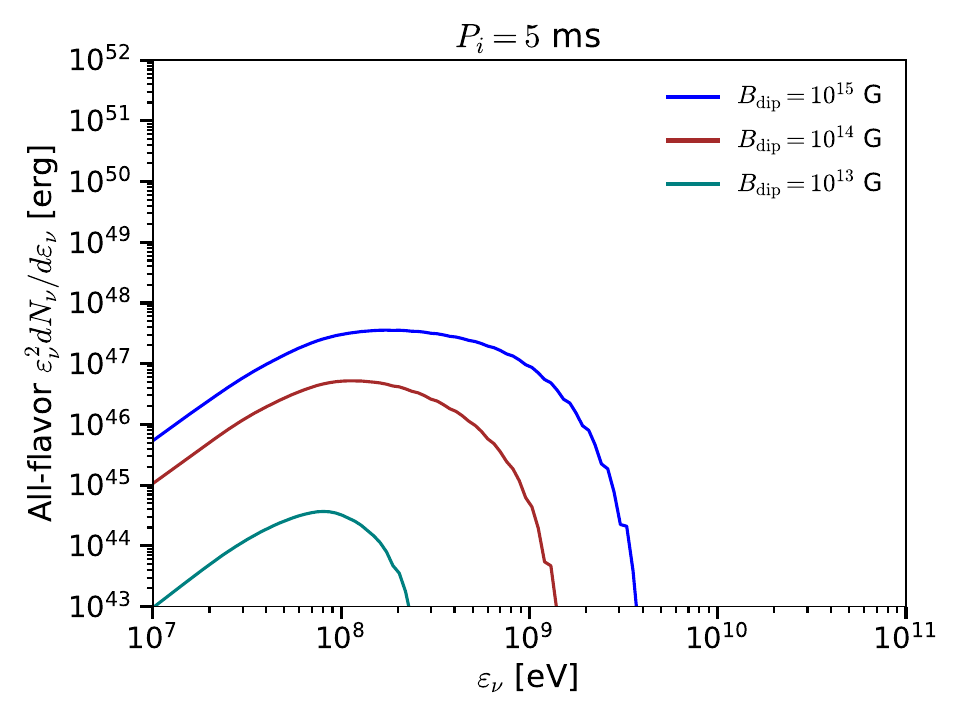}
    \caption{Time-integrated all-flavor neutrino spectrum $\varepsilon_{\nu}^2 dN_{\nu}/d\varepsilon_{\nu}$, for $Y_n = 0.5$. Left (right) panel corresponds to fixed $B_{\rm dip}=10^{15}\,{\rm G}$ ($P_i = 5\,{\rm ms}$).
    } 
\label{NuSpectrum}
\end{figure*}

For $np$ inelastic collisions and neutrino production to begin, neutron decoupling needs to occur first. As the wind propagates through the SN ejecta, ions in the winds are decelerated at the shock. Then the decoupled neutrons propagate through the wind and interact in the shocked wind. Alternatively, neutrons eventually cross the shock and will collide with the ejecta material, creating pions. Roughly half of the neutron energy is carried by the leading nucleon. Approximately $\sim1/20$ of the neutron kinetic energy is carried by each neutrino produced after a single $np$ inelastic scattering and subsequent pion decay.

To obtain the spectrum of neutrinos that escape the source, we solve the transport equation for both neutrons and neutrinos. Denoting by $dN (\varepsilon,t)/d\varepsilon$ the number of particles with energies between $\varepsilon$ and $\varepsilon+d\varepsilon$ in the PNS rest frame at time $t$, we solve the transport equations for $dN_n/d\varepsilon_n$ and $dN_\nu/d\varepsilon_\nu$, 
\begin{eqnarray}
\frac{\partial}{\partial t}\frac{dN_n}{d\varepsilon_n} &=& - n_{\rm ej}\sigma_{np}c \frac{dN_n}{d\varepsilon_n}
+Q_n(\varepsilon_n,t)\\\nonumber
& & + \int d\varepsilon_n^\prime \delta(\varepsilon_n-\varepsilon_n^\prime) n_{\rm ej}\sigma_{np}c \frac{dN_n}{d\varepsilon_n^\prime}
\\
\frac{d\dot{N}_{\nu_\alpha}}{d\varepsilon_{\nu_\alpha}} &=& n_{\rm ej}c\int_{\varepsilon_{\nu_\alpha}}^\infty\frac{dN_n}{d\varepsilon_n}\frac{d\sigma_{np}}{d\varepsilon_{\nu_\alpha}}(\varepsilon_n,\varepsilon_{\nu_\alpha})d\varepsilon_n.
\label{TransportEqn}
\end{eqnarray}
The neutron source term is given by $Q_n(\varepsilon_n,t) = (Y_n\dot{M}_b/m_n)H(t-t_{\rm dec})\delta(\varepsilon_n-\Gamma_n(t)m_nc^2)$, where $H$ is the step function. This source term corresponds to a monoenergetic spectrum injection rate normalized to $L_n$, and we only consider neutrons injected after $t_{\rm dec}$, when they are able to cross into the SN ejecta. When neutrons at energy $\varepsilon_n$ interact, they are reinjected with half their kinetic energy, that is, $\varepsilon_n^\prime = m_n c^2 +\kappa_{np}(\Gamma_n(t)-1)\varepsilon_n$, where we take $\kappa_{np}=0.5$.
The target proton density $n_{\rm ej}$ is the stellar ejecta density, which is much larger than $n_w$, enabling neutron depletion. Pions are produced via $np$ collisions, so we only consider $\alpha = e,\mu$. In these equations $N_{\nu_\alpha}$ corresponds to the sum of neutrinos and antineutrinos of flavor $\alpha$. 

The cross section $\sigma_{np}$ and the differential cross sections $d\sigma_{np}/d\varepsilon_{\nu_\alpha}$for $np$ collisions are obtained numerically with \textsc{Geant4} by Refs.~\cite{Murase:2013hh,Murase:2022vqf}. The differential cross sections for neutrino production assume that pions decay quickly, which is a good approximation as the decay rate for $\sim 1-10\,{\rm GeV}$ pions is significantly larger than the relevant cooling rates at these energies. Likewise, the neutrino absorption term at these energies is negligible for the target densities in this environment. We propagate until $t=t_{\rm thin}$, when neutron injection stops and all neutrinos produced during $t_{\rm dec}<t<t_{\rm thin}$ propagate freely and escape the source. Contributions from later times are negligible because the neutron luminosity $L_n$ has significantly decreased. The spectrum at $t=t_{\rm thin}$ is also equal to the spectrum at escape, as we can neglect neutrino attenuation at these low neutrino energies.

In Fig.~\ref{NuSpectrum} we show the time-integrated all-flavor neutrino spectra (fluence), for $B_{\rm dip} = 10^{15}\,{\rm G}$ and various $P_i$ on the left panel, and with $P_i=5\,{\rm ms}$ and various $B_{\rm dip}$ in the right panel. From the neutrino spectra, we see that most of the neutron energy is deposited into $\sim 0.1-5\,{\rm GeV}$ neutrinos. We note that the choice of the time window, $\Delta T = t_{\rm thin}$, may affect the spectra based on the value of $t_{\rm dec}$ for different PNS parameters.
If $t_{\rm dec}>t_{\rm thin}$, the fluence is small regardless of $\Delta T$. If $t_{\rm dec}<t_{\rm thin}$, then the fluence has contributions from $t_{\rm dec} < t <\Delta T$, and will saturate for $\Delta T\gtrsim t_{\rm thin}$. Finally, if $t_{\rm dec}\sim t_{\rm thin}$, we need $t_{\rm thin}<\Delta T$ to obtain a nonnegligible fluence.
The spectra shown in Fig.~\ref{NuSpectrum} do not have significant contributions from $t>\Delta T = t_{\rm thin}$, except for $B_{\rm dip}=10^{13}\,{\rm G}$ and $P_i=5\,{\rm ms}$. In this exceptional case, the spectrum drops to zero if the time integral is done only up to $\Delta T= t_{\rm thin}$, because of $t_{\rm dec}>\Delta T$ (see Fig.~\ref{NeutronLuminosity} right panel). 
We discuss different choices of $\Delta T$ in Section \ref{Sec6}.

Each neutrino spectrum has an energy cutoff that reflects the maximum neutrino energy limited by the parent neutron's Lorentz factor at decoupling.
It takes longer to begin neutrino production for weaker $B_{\rm dip}$ and longer $P_i$, because these parameters slow down the growth of $\Gamma_n$, as shown in Fig.~\ref{REvolution}. 

\subsection{Quasithermal neutrino detection}
Neutrinos in the $0.1-10\,{\rm GeV}$ energy range can be observed in detectors such as HK, KM3NeT-ORCA and IceCube-Upgrade. We consider a Galactic SN located at a distance $d=10\,{\rm kpc}$. We convert the spectra $dN_{\nu_\alpha}/d\varepsilon_{\nu_\alpha}$ into an observed flux $\phi_{\nu_\alpha}$ at the observed neutrino energy $E_{\nu_\alpha}$ using
\begin{equation}
\phi_{\nu_\alpha}(E_{\nu_\alpha}) = \frac{1}{4\pi d^2}\sum_{i=1}^3\sum_{\beta = e,\mu,\tau}|U_{\alpha i}|^2|U_{\beta i}|^2 \left.\frac{dN_{\nu_\beta}}{d\varepsilon_{\nu_\beta}}\right|_{\varepsilon_{\nu_\beta}=E_{\nu_\alpha}},
\end{equation}
where the assumption $\varepsilon_{\nu_\beta}=E_{\nu_\alpha}$ holds for sources at negligible redshift, as is the case for a Galactic SNe. We calculate the event rates using 
\begin{equation}
\mathcal{N} = \int dE_\nu A_{\rm eff}(E_\nu) \phi_{\nu}(E_\nu),
\label{NEvts}
\end{equation}
where $A_{\rm eff}$ is the neutrino effective area depending on the neutrino flavor and $\phi_{\nu}$ is the neutrino flux.

To calculate the number of neutrino events in HK, we use the effective areas reported in Ref.~\cite{SK:2021dav} for SK. The events in SK are divided into three classes: fully contained (FC), partially contained (PC) and upward-going muons (UPMU). In the case of FC events, we scale the effective area by the fiducial mass ratios between HK and Super-Kamiokande (SK), which is roughly 8.3
(187 kton for HK and 22.5 kton for SK). The scaling for PC events is not straightforward because the events initiate outside the detector volume, so we do not consider them in this work. For UPMU events, we use half the effective area reported in Ref.~\cite{Abe:2011ts}, to account for our use of one tank.

In the case of IceCube-Upgrade, the effective areas for all neutrino flavors are taken from Ref.~\cite{Clark:2015idz}. It should be noted that the $\nu_\tau$ effective area is lower than $\nu_e$ and $\nu_\mu$ areas at all energies. For KM3Net-ORCA, the effective volumes $V_{\rm eff}$ for $\nu_e$ and $\nu_\mu$ charged current (CC) interactions are available, which can be converted to $A_{\rm eff}$ using
\begin{equation}
    A_{\rm eff} = \sigma^{\rm CC}\rho N_A V_{\rm eff},
\end{equation}
where $\sigma^{\rm CC}$ is the CC interaction cross section, $N_A$ is Avogadro's constant and $\rho$ is the water density.

The neutrino energy threshold in HK is $100\,{\rm MeV}$ for FC events. IceCube-Upgrade and KM3Net-ORCA have energy thresholds of $\approx 3\,{\rm GeV}$. Within the $0.1-10\,{\rm GeV}$ neutrino energy range, the main background is due to atmospheric neutrinos. We adopt the atmospheric neutrino flux model by Honda et al. (2011)~\citep{HKKM11} (henceforth, the HKKM flux) to calculate the atmospheric background rates in all three detectors (see Ref.~\cite{Super-Kamiokande:2015qek} for the comparison with SK atmospheric neutrino flux measurements). We use the angle-averaged HKKM flux in Fig. 7 of Ref.~\cite{Super-Kamiokande:2015qek}. We address the impacts of solar modulation in Section~\ref{Sec6}.

\begin{table}
\centering
\def\arraystretch{1.4}
\caption{Expected number of $E_\nu > 100\,{\rm MeV}$ events for a source located at $d=10\, {\rm kpc}$, evaluated for various PNS configurations within the chosen time window ($t<\Delta T$). We show the event numbers for KM3Net-ORCA and IceCube-Upgrade for CC interactions and the sum of FC and UPMU events in HK.
}
\scalebox{1.0}{\begin{tabular}{|c|c|c|c|}
\hline
$(B_{\rm dip}/{\rm G},P_i/{\rm ms})$ &  HK & KM3Net- & IceCube- \\
& & ORCA & Upgrade \\
\hline
($10^{15}$, $1$)  & $2.1\times 10^4$ & $1.4\times 10^4$ & $4.6\times 10^4$\\
($10^{15}$, $3$)  & $1.2\times10^2$ & 13 & $89$\\
($10^{15}$, $10$)  & 1.8 & $1.2\times10^{-2}$ & $5.1\times10^{-2}$\\
($10^{15}$, $30$)  & $2.1\times 10^{-2}$ & $\sim 0 $ & $\sim$ 0\\ \hline
($10^{14}$, $1$)  & $9.8\times10^{2}$ & $2.0\times10^{2}$ & $1.1\times10^{3}$\\
($10^{14}$, $3$)  & 8.0 & $1.1\times10^{-1}$ & $7.7\times 10^{-1}$\\
($10^{14}$, $10$)  & $2.7\times 10^{-2}$ & $\sim 0$ & $\sim 0$\\
($10^{14}$, $30$)  & $\sim 0$ & $\sim 0$ & $\sim 0$\\ \hline
($10^{13}$, $1$)  & $4.5\times10^{-1}$ & $\sim 0$ & $\sim 0$\\
($10^{13}$, $3$)  & $2.2\times 10^{-2}$ & $\sim 0$ & $\sim 0$\\
($10^{13}$, $10$)  & $\sim 0$ & $\sim 0$ & $\sim 0$\\
($10^{13}$, $30$)  & $\sim 0$ & $\sim 0$ & $\sim 0$ \\
\hline
\end{tabular}}
\label{NuEventsTable}
\end{table}

Based on our choice $\Delta T = t_{\rm thin}$, we multiply the HKKM flux by $t_{\rm thin} = 70\,{\rm s}$ to obtain a time-integrated background flux which can be integrated over solid angle and be inserted in Eq.~\eqref{NEvts} to estimate the number of background events.
We find 0.27, 0.43 and 0.77 events per 70 seconds in HK, KM3Net-ORCA and IceCube-Upgrade, respectively. 
The total number of neutrino events with energies $E_\nu>100\,{\rm MeV}$ for each of the three detectors is summarized in Table~\ref{NuEventsTable}. The biggest difference between different PNS parameters is the neutron luminosity, which is also reflected in the number of events. 
\begin{figure}
    \centering
\includegraphics[width=1.02\columnwidth]{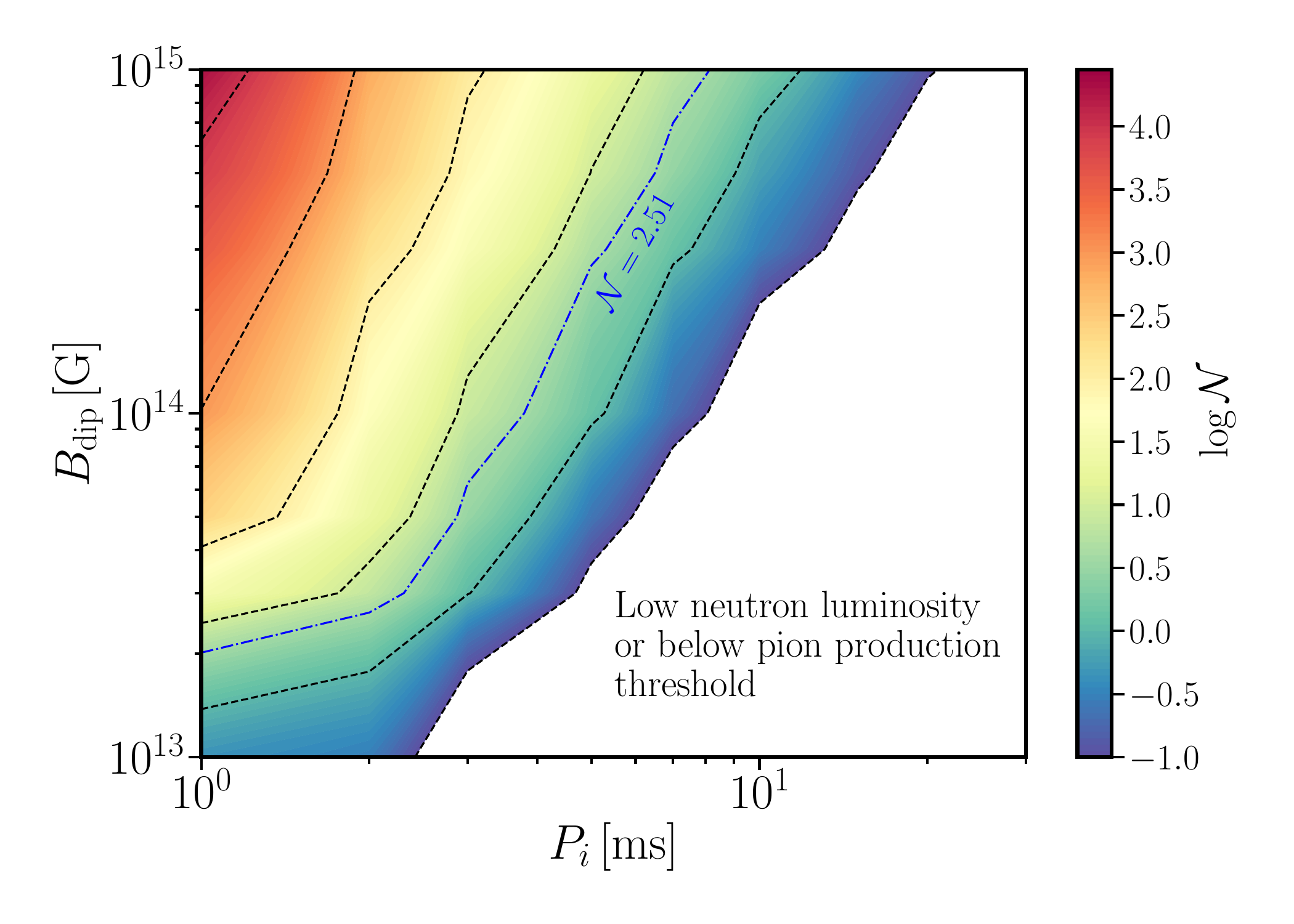}
    \vspace{-0.6cm}
    \caption{Number of events with $E_\nu>100$ MeV over the $B_{\rm dip} - P_i$ parameter space in HK, for a search window $\Delta T = t_{\rm thin}$. The source distance is $d=10$ kpc. The sensitivity of $\mathcal{N} = 2.51$, calculated via the Feldman-Cousins approach, is shown as a dashed blue curve.}
    \label{EventContours}
\end{figure}

In Fig.~\ref{EventContours} we show the number of events in HK. The corresponding number of background events is 0.27. From this background we can derive the sensitivity, defined as the average $90\%$ CL Feldman-Cousins upper limit \cite{Feldman:1997qc}. For HK, this limit is $\mathcal{N}<2.51$, represented by the dashed blue curve. The empty white region in Fig.~\ref{EventContours} corresponds to $\mathcal{N}<0.1$. Here, the pion production threshold is reached close to $t_{\rm thin}$ or later, when $L_n$ is already low. There are also cases for which the threshold is not reached before neutron decoupling takes place, resulting in no quasithermal neutrinos (see Table~\ref{NuEventsTable}).

Finally, we note that our numerical calculations confirm the analytical estimates presented in MDT14. The difference in the neutrino fluence is mostly due to $f_{\rm op}$ and $f_{\rm cen}$. For example, we have $f_{\rm cen}\approx1.6$ and $f_{\rm op}\approx6.9\times{10}^{-2}$ for $P_i=5$~ms, , and $f_{\rm cen}\approx1.2$ and $f_{\rm op}\approx4.1\times{10}^{-2}$  for $P_i=10$~ms. The coasting neutron Lorentz factor is also affected by $f_{\rm op}$. Also, for HK, MDT14 assumes 0.56~Mt while we use 0.187~Mt.

\section{Discussion} 
\label{Sec6}
In Section~\ref{Sec4}, we have shown that nuclei can be photodisintegrated by nonthermal photons for our chosen PNS parameter space. In this case the role of $Y_e$ is to act as a normalization factor for $Y_n = 1-Y_e$ as it becomes a time-independent quantity during the neutrino emission.

The neutrino fluxes in Fig.~\ref{NuSpectrum} were obtained for $t_{\rm thin}=70\,{\rm s}$ and a stretch factor $\eta_s = 3$. The stretch factors relate the neutrino cooling curve parameters of rapidly rotating and non-rotating PNS using \cite{Metzger2011a}
\begin{eqnarray}
L_\nu \to L_{\nu,\Omega=0}\eta_s^{-1};
t\to t_{\Omega=0}\eta_s;
\varepsilon_\nu\to\varepsilon_{\nu,\Omega=0}\eta_s^{-1/4}.
\end{eqnarray}
The stretch factor $\eta_s$ is expected to be an increasing function of the PNS angular velocity $\Omega$. As a result, we should have $t_{\rm thin}\propto \eta_s \propto \Omega$, although we assume a fixed value for $\eta_s$ in our analysis Therefore, it is possible that $t_{\rm thin}$ is somewhat underestimated (overestimated) for short (long) $P_i$. The modifications to $L_\nu$ will also affect the neutron luminosity $L_n$.

The production of GeV neutrinos has been extensively discussed in the context of GRBs (e.g., Refs.~\cite{Derishev1999,Bahcall:2000sa,Koers:2007ww,Beloborodov:2009be,Murase:2013hh}). We stress that the main neutrino production mechanism of our scenario is different from that in GRBs. In GRBs, there are three possibilities. First, while neutrons decouple, protons can continue to accelerate and their relative motion may be sufficient to enable high-energy $np$ collisions within the outflow (e.g., \cite{Bahcall:2000sa,Koers:2007ww,Murase:2022vqf}). 
Second, after the decoupling, faster proton outflows may catch up with slower neutron outflows, inducing internal $pn$ collisions~\cite{Beloborodov:2009be,Murase:2013hh}. Alternatively, $pn$ and $pp$ collisions can happen between compound flows after the decoupling~\cite{Meszaros:2000fs,Murase:2013hh}. In such a collision model, all neutrons do not have to be dissipated and the situation depends on the $pn$ optical depth~
\cite{Murase:2022vqf}. On the other hand, in our scenario~\cite{Murase:2014bfa}, all decoupled neutrons with Lorentz factors of $\Gamma_n\gtrsim 1.37$ must dissipate via inelastic collisions with nonrelativistic protons, and the system is always regarded as an ideal beam dump. Note that neutrinos produced during the $np$ decoupling are typically subdominant. The typical optical depth for $np$ collisions during the decoupling is $\mathcal{O}(10^{-2})$ for $Y_n\sim 1$ and scales with $Y_n^{1/2}$~\cite{Koers:2007ww}. The $np$ optical depth in our scenario is proportional to $Y_n$, so neutrino contributions from the decoupling can be comparable only when $Y_n\ll 1$.

Regardless of the situation leading to inelastic $np$ collisions, neutral pions are inevitably produced as a result of these interactions, with GeV gamma-ray emission. However, in our scenario, this signal must be attenuated as they propagate through the SN ejecta. The Bethe-Heitler optical depth is very large at times $t\lesssim 10^5~{\rm s}$ and GeV-TeV gamma rays cannot leave the ejecta until $t\gtrsim 50~{\rm days}$ post core-collapse, due to two-photon annihilation \cite{Murase:2014bfa}.

{Note that, by using the angle-averaged HKKM flux in Ref.~\cite{Super-Kamiokande:2015qek}, our calculation of the number of UPMU events does not account for the difference between upgoing and downgoing atmospheric neutrino fluxes.
For SK in particular (and HK by extension), which has a low energy threshold, solar modulation can affect the atmospheric flux normalization by $\sim 20\%$ for downgoing neutrinos below 1~GeV~\cite{Super-Kamiokande:2015qek}. For the other two detectors, solar modulation does not affect neutrinos at GeV energies. We point out that a $\sim 20\%$ difference in the atmospheric neutrino flux for HK would still lead to less than 1 fully contained event within our chosen time window. Applying these corrections to the atmospheric flux is not expected to result in a significant difference in the background rates, so our statements regarding detectability remain unchanged.

We find that HK is sensitive to neutrino emission from hidden winds in PNSs. When $B_{\rm dip}=10^{15}$ G, HK is sensitive to neutrino emission from $P_i<7$ ms. As the PNS magnetic field gets weaker, the sensitivity to $P_i$ weakens, dropping to $P_i<3.2$ ms for $B_{\rm dip}=10^{14}$~G. For weak fields $B_{\rm dip}\sim 10^{13}$~G, neutrino events are significantly reduced as it takes very long for the outflow to decouple, and the Lorentz factors are not significantly above the threshold value of 1.37. In Fig.~\ref{NuSpectrum} we see that there is a flux of $1-10$~GeV neutrinos, which can potentially be detected by IceCube-Upgrade and KM3Net-ORCA. However, once we estimate the number of events from these detectors, the sensitivity only improves marginally. This is due to the fact that an increase in the number of signal events is offset by the increased background rate from the detectors. Consequently, the combined sensitivity curve corresponding to $\mathcal{N}<3.85$ almost overlaps completely with the blue curve depicted in Fig.~\ref{EventContours}, meaning that the parameter space sensitivity remains essentially the same. The calculation of $\mathcal{N}$ also implicitly assumes that the explosion timing is known. Estimates of the explosion time can be made via neutrino - gravitational wave \cite{Bartos:2012vd} or neutrino - optical~\cite{2010APh....33...19C} coincident detections. If the uncertainty on the time window estimate is $\gtrsim 10\;{\rm s}$, then the atmospheric background can be larger, which in turn would weaken our sensitivity.

The choice of $\Delta T = t_{\rm thin}$ is motivated by the behavior of $\dot{M}_b$ as a function of time, such that the neutron luminosity would significantly decrease after $t_{\rm thin}$ when compared to earlier times. Extending the signal search time window to $\Delta T = 100\,{\rm s}$ only leads to $\lesssim 1\%$ differences in the time-integrated spectra for our most energetic configurations. In this case, our overall sensitivity also decreases as we accumulate more background events and the signal-to-noise ratio decreases. The effect of shortening $\Delta T$ depends on the PNS parameters, specifically their impact on $t_{\rm dec}$. Decoupling can take up to $80\,{\rm s}$, with longer $P_i$ and weaker $B_{\rm dip}$ typically leading to longer $t_{\rm dec}$. If $t_{\rm dec}\sim t_{\rm thin}$, then reducing $\Delta T$ has a major impact on the neutrino signal. The accompanying MeV neutrino signal from the SN would be detected in HK, and help us estimate $t_{\rm thin}$, which can then be used as the time window $\Delta T$ for the GeV neutrino signal search. Otherwise, $\Delta T\approx 70-100\;{\rm s}$ would be a reasonable choice for observational search.

It should be noted that inclusion of more neutrino detectors does not significantly improve our access to the white region in Fig.~\ref{EventContours}, where neutrino emission from the source itself is significantly suppressed. While detection prospects are not improved by using additional detectors, the increased statistics in the parameter space that we are sensitive to can be used to narrow down the $(B_{\rm dip},P_i)$ parameter space for PNS progenitors, once these high-energy neutrinos are detected.

\section{Summary}
\label{Sec7}
The majority of core-collapse SNe are believed to leave PNSs as compact remnants, and some of them may turn out to be rapidly rotating and/or strongly magnetized. The PNS wind eventually becomes Poynting dominated, and their magnetic energy will be converted into the bulk kinetic energy. Given that free neutrons are entrained in the wind, inelastic $np$ collisions lead to pions that decay into neutrinos in the $0.1-10\,{\rm GeV}$ range. The neutrino energy at which $\varepsilon_\nu^2 dN_\nu/d\varepsilon_\nu$ peaks is insensitive to the PNS parameters and most of the neutrinos are emitted within the first $\sim 70$~s post-collapse. While we confirmed the scenario by MDT14, our results are more quantitative. 
The mechanism considered in this work relies on bulk acceleration of the outflow but not on cosmic-ray acceleration.  

Our results pave the way to study the next Galactic SN as a multienergy neutrino source from MeV to PeV energies. Using the MeV thermal neutrino component from the CCSN would also be useful in removing parameter degeneracies. The MeV neutrinos contain information on $L_\nu$ from the neutrino cooling phase. This would allow us to separate $L_\nu$ from $f_{\rm op}$, $f_{\rm cen}$ and $\Gamma_n$ which are related to the magnetization. 
Detection of quasithermal GeV neutrinos is a strong indicator of relativistic neutron outflows. The next Galactic SN may give us an opportunity to study PNS parameters through quasithermal neutrinos. Rapidly rotating magnetars are particularly motivated by observations of SNe Ibc, including SNe Ibc-BL and SLSNe-I, in which 
nonthermal neutrinos with $E_\nu\gtrsim 1\,{\rm TeV}$ and accompanying electromagnetic emissions may also be expected \cite{Murase:2009pg,Kashiyama:2013ata,Murase:2017pfe}. Multienergy neutrino emission can help us understand the connection between SNe and GRBs.\\

\begin{acknowledgments}
J.C. was supported by the NSF Grant No.~AST-1908689 and No.~AST-2108466. J.C. also acknowledges the Nevada Center for Astrophysics and NASA award 80NSSC23M0104 for support. N.E. is supported by NSF Grant No.~AST-1908960. M.B. acknowledges support from the Eberly Research Fellowship at the Pennsylvania State University and the Simons Collaboration on Extreme Electrodynamics of Compact Sources (SCEECS) Postdoctoral Fellowship at Wisconsin IceCube Particle Astrophysics Center (WIPAC), University of Wisconsin-Madison. The work of K.M. is supported by the NSF Grant No.~AST-1908689, No.~AST-2108466, No.~AST-2108467, and No.~AST-2308021, and KAKENHI No.~20H01901 and No.~20H05852. The work of S.H.~is supported by the U.S.~Department of Energy Office of Science under award number DE-SC0020262, NSF Grant No.~AST1908960 and No.~PHY-2209420, JSPS KAKENHI Grant Number JP22K03630 and JP23H04899, and the Julian Schwinger Foundation. This work was supported by World Premier International Research Center Initiative (WPI Initiative), MEXT, Japan.
\end{acknowledgments}

\bibliography{main}

\label{lastpage}
\end{document}